\pdfoutput=1
\documentclass[onecolumn,binary-units=true,superscriptaddress]{revtex4}
\usepackage[T1]{fontenc}
\usepackage[utf8]{inputenc}

\usepackage[english]{babel}
\usepackage{amsfonts, amsmath, amssymb, amsthm, nicefrac}
\usepackage{siunitx}
\usepackage{array,booktabs}
\usepackage[tableposition=top]{caption}
\usepackage{subcaption}
\usepackage{csquotes}
\usepackage{multirow}
\usepackage{enumitem}
\usepackage{siunitx}
\usepackage{bbm}
\usepackage{tikz}
\usepackage[textsize=small]{todonotes}
\usepackage{hyperref}
\usepackage{soul}

\usetikzlibrary{arrows.meta, calc, matrix, positioning, shapes, fit, graphs, quotes}

\newtheorem{defn}{Definition}
\newtheorem{lem}{Lemma}
\newtheorem{prop}{Proposition}

\newtheorem{example}{Example}
\theoremstyle{remark}
\newtheorem{rem}{Remark}

\makeatletter
\let\oldtheequation\theequation
\renewcommand\tagform@[1]{\maketag@@@{\ignorespaces#1\unskip\@@italiccorr}}
\renewcommand\theequation{(\oldtheequation)}
\makeatother
\AtBeginDocument{%

}

\usepackage{algorithm}
\usepackage[noend]{algpseudocode}

\usepackage{natbib}
\usepackage{graphicx}

\usepackage{braket}

\def\NN{\mathbb{N}}

\def\RR{\mathbb{R}}
\def\CC{\mathbb{C}}
\def\EE{\mathbb{E}} %
\def\Dcal{\mathcal{D}}
\def\Hcal{\mathcal{H}}
\def\Pcal{\mathcal{P}}

\def\Rcal{\mathcal{R}}

\def\indicator{\mathbbm{1}}

\DeclareMathOperator{\Tr}{Tr}
\DeclareMathOperator{\supp}{supp}

\DeclareMathOperator{\Var}{Var}
\DeclareMathOperator{\Lift}{Cover}

\newcommand{\pt}[1]{\ensuremath{\mathit{#1}}}

\begin{document}

\title{Decision Diagrams for Quantum Measurements with Shallow Circuits}

\author{Stefan Hillmich}
\email[Corresponding author: ]{stefan.hillmich@jku.at}
\affiliation{Johannes Kepler University Linz, 4040 Linz, Austria}
\author{Charles Hadfield}
\email[Corresponding author: ]{charles.hadfield@ibm.com}
\affiliation{IBM Quantum, IBM T.J. Watson Research Center, Yorktown Heights, NY 10598}
\author{Rudy Raymond}
\email[Corresponding author: ]{rudyhar@jp.ibm.com}
\affiliation{IBM Quantum, IBM Japan, 19-21 Nihonbashi Chuo-ku, Tokyo, 103-8510, Japan}
\affiliation{Quantum Computing Center, Keio University, 3-14-1 Hiyoshi, Kohoku-ku, Yokohama, Kanagawa, 223-8522, Japan}
\author{Antonio Mezzacapo}
\affiliation{IBM Quantum, IBM T.J. Watson Research Center, Yorktown Heights, NY 10598}
\author{Robert Wille}
\affiliation{Johannes Kepler University Linz, 4040 Linz, Austria}
\affiliation{Software Competence Center Hagenberg (SCCH) GmbH, 4232 Hagenberg, Austria}

\begin{abstract}
    We consider the problem of estimating quantum observables on a collection of qubits, given as a linear combination of Pauli operators, with shallow quantum circuits consisting of single-qubit rotations. We introduce estimators based on randomised measurements, which use decision diagrams to sample from probability distributions on measurement bases. This approach generalises previously known uniform and locally-biased randomised estimators. The decision diagrams are constructed given target quantum operators and can be optimised considering different strategies. We show numerically that the estimators introduced here can produce more precise estimates on some quantum chemistry Hamiltonians, compared to previously known randomised protocols and Pauli grouping methods.   
\end{abstract}
\maketitle
\section{Introduction}

Variational quantum algorithms are based on a quantum-classical optimisation feedback loop, in which a trial parameterised quantum state is prepared on a quantum computer, a target quantum cost function is estimated on it, and a classical optimiser changes the quantum parameters to minimise the target observable. This machinery has been instrumental to find ground state energies of quantum chemistry systems, which are the smallest quantum systems believed to deliver quantum advantage in the field of quantum simulations. 

Recent experiments on variational quantum  algorithms~\cite{kandala2017hardware,o2016scalable,hempel2018quantum} have shown that precise estimates of complex quantum operators are essential for a successful execution of the algorithms. A finite single-qubit measurement budget can hinder the performance of the quantum-classical optimisation cycle, by introducing stochastic noise on the quantum cost function to be optimised.
This problem is particularly severe for quantum chemistry systems, whose molecular Hamiltonians are composed of a linear combination of Pauli operators that grows, at worst, with the fourth power of the system size~\cite{Wecker2015ProgressAlgorithms}.

To alleviate the measurement problem, a variety of algorithms have been proposed. They all ultimately aim at obtaining precise estimations of multi-qubit quantum operators, typically given as linear combination of Pauli operators, with the smallest amount of single-qubit measurements.

The idea of using the same single-qubit measurements to estimate grouped Pauli operators that qubit-wise commute, introduced in~\cite{kandala2017hardware} as \emph{tensor-product basis sets}, is at the core of several measurements protocols. Some of them promise a reduction in the number of measurements for quantum chemistry systems, exploiting Pauli grouping heuristics, at the expense of an increase depth in the quantum circuits used to prepare the state to be measured~\cite{gokhale2019minimizing,izmaylov2019unitary,crawford2019efficient}. However increased circuit depths can impair the execution on noisy quantum computers prone to decoherence. Furthermore, even in the fault-tolerant regime, bigger circuit depths can increase the overall runtime of the quantum algorithm.
Addressing molecular systems, \cite{zhao2019measurement} shows a linear saving in the number of grouped Pauli operators when addressing molecular systems through unitary partitioning of a target Hamiltonian, while ~\cite{huggins2019efficient} finds a cubic reduction if the problem is expressed in plane wave basis, at the expense of a linear increase in circuit depth. \cite{crawford2019efficient} proposes \textit{sorted insertion} to group Pauli operators based on their weights, preprocessing computations linear in the number of qubits and quadratic in the number of Pauli operators. Refs. \cite{gokhale2019minimizing} and \cite{hamamura2019efficient} exploit simultaneous measurability of partitions of commuting Pauli strings. Exploiting the automated search for symmetries introduced \cite{bravyi2017tapering}, \cite{yen2019measuring} shows a linear scaling when applied to chemistry problems, again at the expense of increased circuit depth. \cite{jena2019pauli} considers random Pauli sets, and uses greedy graph coloring algorithms to determine partition of Pauli operators of Hamiltonians,  conjecturing a linear saving in number of measurements if arbitrary Clifford operators before measurement are allowed. 

Other approaches address the measurement problem while not increasing circuit depths. We refer to this specific case as the \emph{shallow-circuit measurement problem}. It has been tackled so far formulating it in terms of graph coloring, which was solved with a variety of heuristics~\cite{verteletskyi2020measurement,yen2019measuring}.
Hybrid architectures made of quantum computers in conjunction with trained neural-network quantum states have been employed to reduce measurement variances~\cite{torlai2019precise}.

The results contained in this work build on recent techniques based on randomised sequences of single-qubit measurements. A framework for efficiently estimating properties of reduced subsystems of quantum states was introduced in~\cite{Huang2020PredictingMeasurements}. There, collections of randomised measurement outcomes, labeled \emph{classical shadows}~\cite{aaronson2019shadow}, are classically stored to retrieve at a later stage expectation values of local observables. While this procedure is very well suited for retrieving many generic local observables, the uniform distribution used to draw measurement bases is not optimal in estimating with high precision specific observables such as molecular Hamiltonians. 
Building on this result, \cite{huang2021efficient} uses \emph{derandomisation} to deterministically change a sequence of measurement bases drawn at random, with the goal of improving precision in estimating specific sets of Pauli operators.

Improving on uniform distribution sampling, biased randomised measurement protocols can be used to improve estimate precision of given observables. The \emph{locally-biased classical shadows} introduced in~\cite{hadfield2020} are collections of random measurements generated by probability distributions optimised locally at the single-qubit level. While a bias on single-qubit product probability distributions can outperform on uniform random distributions and Pauli grouping heuristics, it still misses on improvements that can come by considering generic measurement probability distributions on a set of qubits.       
Here we introduce a framework to sample from probability distributions on measurement bases that generalises the local product probability distributions considered in~\cite{hadfield2020}. These probability distributions are generated using decision diagrams, constructed from a target quantum observable, given as a linear combination of Pauli operators. 

Decision diagrams are a well-known graph-based data structure used in many disciplines of computer~science to enable compact representation of data in many cases.
Example applications include binary decision diagrams representing Boolean functions~\cite{DBLP:conf/dac/Bryant85}, zero-suppressed binary decision diagrams with a focus on sets~\cite{DBLP:journals/sttt/Minato01}, tagged binary decision diagrams as a combination of both~\cite{DBLP:conf/fmcad/DijkWM17}, $\pi$DDs representing permutations~\cite{DBLP:conf/sat/Minato11}, as well as decision diagrams representing quantum states and quantum operations~\cite{DBLP:conf/date/AbdollahiP06,DBLP:journals/ieicet/WangLTK08,DBLP:books/daglib/0027785,DBLP:journals/tcad/NiemannWMTD16,DBLP:conf/iccad/ZulehnerHW19}.
At their core, decision diagrams decompose the given data into smaller parts by successively making decisions to remove degrees of freedom, recording these decisions, and exploiting the emergence of parts that are equal.
For probability distributions for the measurement problem, decision diagrams provide a natural way to bias the selection of the next measurement basis based on the previous decisions. Decision diagrams have 
been used for efficient sampling on large sets by applying dynamic programming methods~\cite{pmlr-v84-sakaue18a}. 

We propose and discuss different strategies to build decision diagrams and to optimise them in order to reduce the variance of quantum observable estimators that rely on them. We show numerically that they can improve estimation precision on some molecular Hamiltonians, compared to locally biased probability distributions.
The implementation we base the results on is available \url{https://github.com/iic-jku/dd-quantum-measurements}.

\subsection*{Outline of the paper}
We introduce the shallow-circuit measurement problem in \autoref{sec:problem_definition} together with the general probabilistic measurement framework, reviewing the locally-biased approach. 
We define the decision diagrams and present the main idea in \autoref{sec:DD estimators}, followed by strategies to construct decision diagrams given target Hamiltonian operators in \autoref{sec:constructing-DD}.
In \autoref{sec:experimental-evalutation}, we numerically benchmark the estimators bases on decision diagrams versus existing approaches, on quantum Hamiltonians of increasing sizes, representing quantum chemistry models. We summarize the results obtained and draw conclusions in \autoref{sec:conclusion}. Some technical details on the variances of the estimation by decision diagrams, their implementation, optimisation, and relation with previous work are given in Appendix. 

\section{The shallow-circuit Measurement Problem}
\label{sec:problem_definition}

This section precisely establishes the problem of shallow-circuit measurement and a general framework of its solutions as search problems 
over probability distributions. The general framework allows us not only to describe the existing probabilistic solutions, such as, the Pauli grouping via graph coloring and the (locally-biased) classical shadows, but also to 
derive the new probabilistic solution using decision diagrams that overcome the drawbacks of existing solutions. This requires us to introduce some notation which we do progressively in the following subsection.

\subsection{Problem Definition}

Consider an $n$-qubit Hamiltonian
\begin{equation}
    H = \sum_{P \in \{I,X,Y,Z\}^n}
            \alpha_P P \label{eqn:hamiltonian}
\end{equation}
with $\mbox{poly}(n)$ number of real coefficients $\alpha_P$ acting on a quantum processor, we say $P$ is a \emph{Pauli operator} consisting of $n$ \emph{single-qubit} Pauli operators and write $P=\left(\otimes_{i\in[n]}P_i\right)\in\{I,X,Y,Z\}^n$ where $I, X, Y, Z$ are $2\times 2$ Pauli matrices. 
The Hilbert space is $\Hcal :=(\CC^2)^{\otimes n} = \CC^{2^n}$. 
Let $\Dcal(\Hcal)$ denote the space of quantum densities and fix some unknown $\rho\in\Dcal(\Hcal)$. Our task is to estimate $\Tr(H \rho)$ to some additive accuracy $\varepsilon>0$.

We restrict our attention to algorithms for $\Tr(H\rho)$ which are compatible with quantum processors of the current generation (see \autoref{rem:ftqc}). Specifically we assume that the measurement bases in which we may measure $\rho$ are of the form $B=\otimes_{i\in[n]} B_i$ where $B_i=x_i X + y_i Y + z_i Z$ and $x_i^2+y_i^2+z_i^2=1$.
If we then prepare $\rho$ many times, say $S\in\NN$, and for each $s\in[S]$ choose a measurement basis $B^{(s)}$ in which to measure $\rho$ we can estimate, with progressively increasing accuracy, the value of $\Tr(H\rho)$. 
We will make two further assumptions: the choice of $B^{(s)}$ is independent of $B^{(s')}$ for $s'<s$;
any such basis $B$ is a Pauli operator $B\in\Pcal^n$ where $\Pcal=\{X,Y,Z\}$. The \emph{shallow-circuit measurement problem} is how to best choose the measurement bases in order to estimate $\Tr(H\rho)$ within accuracy $\varepsilon$ with as few preparations of $\rho$ as possible.

\begin{rem}\label{rem:ftqc}
Quantum computers of the current generation are not fault-tolerant. This naturally imposes restrictions on the type of measurement schedules we should consider. The precision with which we can implement non-entangling gates (single-qubit unitaries) is significantly greater than that of entangling gates (multi-qubit unitaries). It is therefore natural to consider the restriction of the problem of estimating $\Tr(H\rho)$ to the constrained setting of measuring $\rho$ in bases which do not require entangling gates. In general, reducing entangling gates is beneficial in 
computational time even for fault-tolerant quantum computers. 
\end{rem}
\begin{rem}
The two further assumptions have been chosen to ease the exposition.
The first assumption, that the choice of a basis is independent of previous choices is important. It prevents the situation where a potentially expensive classical computation is performed between each shot. This is similar to the distinction between adaptive and non-adaptive scheduling in annealing schedules.
The second assumption, that measurement bases are full-weight Pauli operators, allows us, in the following section, to talk about probability distributions which are over finite sets. This assumption should be considered minor.
\end{rem}

\subsection{The General Probabilistic Measurement Framework}

We can solve the shallow-circuit measurement problem by viewing it as the problem of how to best pick a probability distribution $\beta$ over the measurement bases $\Pcal^n$. In order to see the relationship we fix some notation.
First, for a fixed Pauli operator $P$, let
\begin{equation}
    \Lift(P) 
    :=
    \left\{
        B \in \Pcal^n
        \,|\,
        B_i = P_i \textrm{ whenever } P_i\neq I
    \right\}.
\end{equation}
This is the set of measurement bases which allow us to estimate $\Tr(P \rho)$. (We shall say that any such $B$ \emph{covers} $P$.)
Next, if $B=\otimes_{i\in[n]} B_i\in\Pcal^n$, then measuring qubit $i$ in the $B_i$ basis returns an eigenvalue $\mu(B,i)\in\{\pm1\}$. 
For a subset $A\subseteq[n]$ let us declare
$\mu(B,A) := \prod_{i\in A} \mu(P,i)$
with the convention that $\mu(P,\varnothing)=1$. If we set $\supp(P):=\{i\in[n] | P_i\neq I\}$ then we find that $\Tr(P\rho)$ is estimated by $\mu(B,\supp(P))$ whenever $B$ covers $P$.
Penultimately, let $\indicator_\Omega$ represent the indicator function of a set $\Omega$. That is $\indicator_\Omega(x)$ returns $1$ if $x\in\Omega$ and $0$ if $x\not\in\Omega$. 
Finally, if $\beta:\Pcal^n\to\RR^+$ is a probability distribution then the probability that a basis $B$ is chosen such that $\Tr(P\rho)$ may be estimated is
\begin{equation}
    \zeta(P,\beta)
    = \sum_{B \in \Pcal^n} \indicator_{\Lift(P)}(B) \cdot \beta(B)
    = \sum_{B\in\Lift(P)}\beta(B).
\end{equation}

\begin{algorithm}[tb]
	\caption{Shallow-measurement estimation of $\Tr(H\rho)$ given $\beta$.}
	\label{alg:general}
	\begin{algorithmic}
	\For{shot $s\in[S]$}
			\State Prepare $\rho$
			\State Select basis $B\in\Pcal^n$ from $\beta$-distribution
			\For{qubit $i\in[n]$}
				\State Measure qubit $i$ in basis $B_i$ giving $\mu(P,i) \in \{\pm 1\}$
			\EndFor
		\State Estimate observable expectation
			\[
			\nu^{(s)} = 
			\sum_P \alpha_P 
			\cdot 
			\frac{ \indicator_{\Lift(P)} (B) }{ \zeta(P,\beta) }
			\cdot
			\mu(B,\supp(P))
			\]
	\EndFor
		\Return $\nu=\frac1S\sum_{s\in[S]} \nu^{(s)}$
	\end{algorithmic}
\end{algorithm}

The shallow-circuit measurement problem reduces to finding $\beta:\Pcal^n\to\RR^+$ which minimises the variance of the estimator $\nu$ produced by \autoref{alg:general}.

Let us say that $\beta$ is \emph{compatible} with the Hamiltonian $H$ if $\zeta(P,\beta)>0$ whenever $\alpha_P\neq 0$. In the Appendix we prove two results: we show that this algorithm returns an unbiased estimator for $\Tr(H\rho)$ provided $\beta$ is compatible with $H$; we also calculate the variance of the estimator. We record these statements below.

\begin{lem}\label{lem:m1}
If $\beta$ is compatible with $H$ then the first moment of the estimator $\nu^{(s)}$ from \autoref{alg:general} is 
\begin{equation}\label{eq:moment1}
	\EE(\nu^{(s)}) = \sum_P \alpha_P \Tr(P \rho).
\end{equation}
\end{lem}

\begin{lem}\label{lem:m2}
If $\beta$ is compatible with $H$ then the second moment of the estimator $\nu^{(s)}$ from \autoref{alg:general} is 
\begin{equation}\label{eq:moment2}
	\EE(\nu^{(s)}\nu^{(s)}) 
	= 
	\sum_{P,Q} \alpha_P \, \alpha_Q \, 
        g(P,Q,\beta)
            \Tr(PQ\rho)
\end{equation}
where
\begin{equation}
    g(P,Q,\beta)
    :=
    \frac1{\zeta(P,\beta)}
    \frac1{\zeta(Q,\beta)}
    \sum_{B\in \Pcal^n} {\indicator_{\Lift(P)}(B)} \cdot {\indicator_{\Lift(Q)}(B)} \cdot \beta(B).
\end{equation}
\end{lem}

\begin{prop}
If $\beta$ is compatible with $H$ then the estimator in \autoref{alg:general} is an unbiased estimator of $\Tr(H\rho)$ and has variance
\begin{equation}
    \Var(\nu)
    =
    \frac1S\left(
    \left(
    \sum_{P,Q} \alpha_P\,\alpha_Q\,
        g(P,Q,\beta)
            \Tr(PQ\rho)
    \right)
    -
    \left(
    \Tr(H\rho)
    \right)^2
    \right).
\end{equation}
\end{prop}

The task is to prescribe $\beta$ in order to obtain a measurement schedule with small variance. In the subsequent sections we show how existing solutions, i.e., the Pauli grouping and (locally-biased) classical shadows can be cast as instances of choosing specific probability distributions and point out their drawbacks. We then show how decision diagrams offer an excellent prescription for prescribing $\beta$.

\begin{rem}
Although we have derived a formula for the variance of the estimator, we would like to point out the role of $\zeta(P,\beta)$ to the variance. Clearly, the square root of the variance (or, the standard deviation of) $\nu^{(s)}$ at \autoref{alg:general} is at most $\max_{B} \sum_P  \indicator_{\Lift(P)} (B) \cdot |\alpha_P|/\zeta(P,\beta)$. Thus, a rule of thumb to minimise the variance is by choosing the distribution $\beta$ so that $\max_{P} 1/\zeta(P, \beta)$ is minimised. This intuition is indeed true as we show 
later for problems of estimating a set of Pauli terms: our inconfidence of the correctness of the estimator is proportional to $\max_{P} 1/\zeta(P, \beta)$. 
\end{rem}

\begin{rem}
How much knowledge of $\rho$ should we allow ourselves in order to prescribe $\beta$? The obvious application of our subroutine is in the context of finding the ground energy of a Hamiltonian using variational quantum eigensolvers. In this context, it is natural to assume we already have an ansatz for the ground state hence we may assume $\rho$ is always \emph{close to} the ground state of the system. Quantifying this phrase is subtle, however in order to illustrate the strength of our proposal, it is not unreasonable to do the following: Prescribe the distribution $\beta$ using \emph{no} knowledge of $\rho$ and then calculate the variance of $\nu$ when given access to the true ground state of the system. Already in this regime, we show significant improvements over existing protocols.
\end{rem}

\begin{rem}
In \hyperref[app:data_structure]{Appendix~\ref*{app:data_structure}} we show how \autoref{alg:general} would be implemented in software.
\end{rem}

\subsection{Existing Solutions and Drawbacks}\label{sec:rel_work}

In the following two subsections,  we review briefly some existing solutions to the shallow measurement problem and their drawbacks that motivate us to develop a new framework with decision diagrams. It is useful to observe that they can be easily seen as instances of \autoref{alg:general}. 

\subsubsection{Pauli Grouping via Graph Coloring}\label{subsec:Pauligrouping}

Pauli grouping via graph coloring has been used experimentally in \cite{kandala2017hardware}. A detailed explanation may be found in \cite[Section A.2]{hadfield2020}. The core idea is to group the Pauli operators $\{P\}_{\alpha_P\neq 0}$ occurring in $H$ into $K$ collections and assign one measurement basis $B^{(k)}\in\Pcal^n$ to each collection $k\in[K]$ such that $B^{(k)}$ allows all Pauli operators in the $k$\textsuperscript{th} collection to be estimated. That is, if $P$ belongs to the $k$\textsuperscript{th} collection then $B^{(k)}\in\Lift(P)$.

The grouping is performed by coloring a specific graph using any graph-coloring heuristic. The graph is constructed first by assigning vertices to each Pauli operator appearing in the Hamiltonian. Second, if two vertices represent Pauli operators $P=\otimes_{i\in[n]} P_i$, and  $Q=\otimes_{i\in[n]}Q_i$, then an edge is added to the graph precisely when there exists a qubit $i\in[n]$ such that $P_i,Q_i\in\Pcal$ and $P_i\neq Q_i$. 
It follows that any vertices with the same color may be assigned a single measurement basis.
Graph coloring heuristics lead to collections for which $K\ll |\Pcal^n|$.

The best assignment of the probability 
$\beta:\{B^{(k)}\}_{k\in[K]}\to\RR^+$
has not been rigorously studied. 
In \cite{kandala2017hardware}, the measurement bases $\{B^{(k)}\}_{k\in[K]}$ are effectively sampled uniformly, but this is due to hardware considerations. 
In \cite[Section A.2]{hadfield2020} an improved sampling is proposed which is based on the $\ell_1$ weight of the coefficients of the Pauli operators appearing in each collection. This proposal may be improved slightly by observing that some Pauli operators may be assigned to several collections. That is, sometimes, several bases from $\{B^{(k)}\}_{k\in[n]}$ may be used to estimate a single Pauli observable. This will decrease slightly the variances obtained in \cite[Tables 1,2]{hadfield2020}.

The main drawback of this approach is observational. That is, on Hamiltonians thus far studied in the literature, the variances are large compared to other proposals. This is despite the preprocessing steps to solve a graph coloring problem to reduce the choices of measurement bases. This may be caused by two reasons. First, the Pauli grouping stage makes no reference to the coefficients $\{\alpha_P\}$. Second, the current proposals may not be optimally assigning the distribution over $\{B^{(k)}\}_{k\in[K]}$.

\subsubsection{Locally-Biased Classical Shadows}\label{subsec:lbcs}
One method of choosing the distribution $\beta:\Pcal^n\to\RR^+$ has been proposed in \cite{hadfield2020} and is called locally-biased classical shadows (LBCS). It may be seen as an extension of a proposal to perform tomography, called classical shadows using random Pauli measurements~\cite{Huang2020PredictingMeasurements}, to the problem of estimating a single observable. LBCS fits precisely into the framework discussed in the preceding subsection. First, the class of distributions from which $\beta$ may be chosen is restricted: only \emph{product} probability distributions are considered, which we may write as $\beta=\prod_{i\in[n]}\beta_i$ where $\beta_i:\Pcal\to\RR^+$ is the probability distribution for choosing to measure the $i$\textsuperscript{th} qubit in a basis $B_i\in\Pcal$. Second, after assuming this restriction on the class of probability distributions, the choice of $\beta$ is made by optimising a convex cost function associated with the Hamiltonian:
\begin{equation}\label{eqn:lbcs_cost}
    \mathrm{cost}_\mathrm{diag}(\beta)
    :=
    \sum_P \alpha_P^2 
        \frac{1}{\prod_{i\in\supp(P)} \beta_i(P_i)}
\end{equation}
A motivation for this choice of cost function can be obtained from \autoref{rem:var_lbcs}. Note that since $\beta$ is assumed to be a product distribution, the denominator $\prod_{i\in\supp(P)}\beta_i(P_i)$ is precisely $\zeta(P,\beta)$. In \cite{hadfield2020}, it was shown that this method leads to significant reductions in the variance of energy estimation in the context of quantum chemistry over the method of Pauli grouping via graph coloring. 

\begin{rem}
Classical shadows, as originally proposed, may be seen as the uniform version of LBCS. That is, $\beta_i(P)=\frac13$ for $P\in\Pcal$. A recent result derandomises this idea and shows promise for estimating single observables~\cite{huang2021efficient}. Derandomisation does however introduce a classical-computation cost because the choice of measurement basis $B^{(s)}$ for some preparation $s\in[S]$ depends on previous shots $s'<s$. Derandomisation should also prove useful to the general framework of decision diagrams. See \hyperref[app:derandomisation]{Appendix~\ref*{app:derandomisation}} where the main result of \cite{huang2021efficient} is generalised to the setting of decision diagrams.
\end{rem}

Both Classical Shadows and LBCS appear to be attractive when Pauli operators in the Hamiltonian are low-weight. That is, when $|\supp(P)|$ is small relative to $n$, since in this case the denominator of \autoref{eqn:lbcs_cost} remains relatively small. This intuition leads to the following setup showing a shortcoming of LBCS. Consider the toy Hamiltonian $H=\otimes_{i\in[n]} X_i + \otimes_{i\in[n]} Z_i$. LBCS would assign, for each qubit, the probability $1/2$ to measure it either in the $X$ or the $Z$ basis. The lack of correlation in these choices implies that the only two bases $X^n$ and $Z^n$ which are useful for energy estimation are rarely chosen. In the parameter $n$ it is with exponentially vanishing probability that such bases are chosen. Pauli grouping via graph coloring would perform much better on this example.
This example motivates the search for distributions $\beta$ from a wider class of probabilities. Although the preceding Hamiltonian is a toy example, there does occur Hamiltonians with similar structure.
For example, for the purpose of \textrm{self-testing} quantum devices, the Hamiltonian %
$H = X_1Z_2\cdots Z_n + \frac1{n-1}\sum_{i\ge 2} Z_1X_i$
is evaluated to test Bell-type inequalities on star graphs~\cite{PhysRevX.7.021042,yang2021testing}.

\section{Estimators of quantum observables with Decision Diagrams}
\label{sec:DD estimators}

This section introduces a new type of decision diagrams. Any such decision diagram (DD) provides a compact representation of a probability distribution over full-weight Pauli operators associated with a given Hamiltonian. To this end, we first briefly review the main idea of decision diagrams in general, before we define and detail the proposed type. Based on that, we show how the new decision diagram can be used to improve the sampling process when measuring quantum Hamiltonians on quantum processors of the current generation. 
Overall, this section provides the motivation and main idea, while details on the technical implementation are covered in the next section.

\subsection{Decision Diagrams in General}

Decision diagrams are a tried and tested data structure in many areas of computer science to provide a compact representation of entities in various domains.
Example applications include binary decision diagrams representing conventional Boolean functions~\cite{DBLP:conf/dac/Bryant85},
zero-suppressed binary decision diagrams with a focus on sets~\cite{DBLP:journals/sttt/Minato01},
tagged binary decision diagrams as a combination of both~\cite{DBLP:conf/fmcad/DijkWM17}, and
$\pi$DDs representing permutations~\cite{DBLP:conf/sat/Minato11}. 
Also in the domain of quantum computing, decision diagrams representing quantum states and quantum operations received interest~\cite{DBLP:conf/date/AbdollahiP06,DBLP:journals/ieicet/WangLTK08,DBLP:books/daglib/0027785,DBLP:journals/tcad/NiemannWMTD16,DBLP:conf/iccad/ZulehnerHW19} and found application, e.g., in the synthesis~\cite{DBLP:conf/date/AbdollahiP06,DBLP:journals/tcad/ZulehnerW18}, simulation~\cite{DBLP:books/daglib/0027785,DBLP:journals/tcad/ZulehnerW19}, or verification~\cite{DBLP:journals/ieicet/WangLTK08,DBLP:journals/tcad/BurgholzerW21} of quantum circuits.

The common idea of all decision diagram-based representations is to decompose a given original representation (e.g., of a Boolean function or quantum state) in a structured fashion that recognizes and exploits redundancies of the decomposed data in order to provide a more compact representation. 
The repeatedly conducted decompositions are represented by means of a directed and acyclic multi-graph, where vertices represent the decomposed data and redundancy is exploited through \emph{shared vertices}.

\begin{figure}
    \centering
    \begin{tikzpicture}[minimum height=0.5cm,minimum width=0.5cm,font=\scriptsize]
    
    \node[draw,circle] (x1) at (1.75,5.5) {\(x_1\)};
    \node[draw,circle] (x2l) at (0.875,4.5) {\(x_2\)};
    \node[draw,circle] (x2r) at (2.625,4.5) {\(x_2\)};
    \node[draw,circle] (x3l) at (0.875,3.5) {\(x_3\)};
    \node[draw,circle] (x3r) at (2.625,3.5) {\(x_3\)};
    \node[draw,circle] (x4l) at (0.875,2.5) {\(x_4\)};
    \node[draw,circle] (x4r) at (2.625,2.5) {\(x_4\)};
    \node[draw] (zero) at (0.875,1.5) {\(0\)};
    \node[draw] (one) at (2.625,1.5) {\(1\)};
    
    \draw (x1) node [above right,xshift=0.25cm] {\(f\)};
    \draw (x2l) node [left,xshift=-0.25cm] {\(f_6=\overline x_2\overline x_3 x_4+x_2x_3\overline x_4\)};
    \draw (x2r) node [right,xshift=0.25cm] {\(f_5=\overline x_2 x_3\overline x_4+x_2\overline x_3 x_4\)};
    \draw (x3l) node [left,xshift=-0.25cm] {\(f_4=x_3\overline x_4\)};
    \draw (x3r) node [right,xshift=0.25cm] {\(f_2=\overline x_3x_4\)};
    \draw (x4l) node [left,xshift=-1cm] {\(f_3=\overline x_4\)};
    \draw (x4r) node [right,xshift=0.25cm] {\(f_1=x_4\)};
    
    \draw (x1) to node[pos=0.25,left] {\tiny 0} (x2l)
    (x1) to node[pos=0.25,right] {\tiny 1} (x2r)
    (x2l) to node[left] {\tiny 1} (x3l)
    (x2l) to node[pos=0.05,right] {\tiny 0} (x3r)
    (x2r) to node[pos=0.05,left] {\tiny 0} (x3l)
    (x2r) to node[right] {\tiny 1} (x3r)
    (x3l) to node[left] {\tiny 1} (x4l);
    \draw (x3l) to [out=225] node[left] {\tiny 0} (zero);
    \draw (x3r) to node[pos=0.25,left] {\tiny 1} (zero)
    (x3r) to node[right] {\tiny 0} (x4r)
    (x4l) to node[left] {\tiny 1} (zero)
    (x4l) to node[pos=-0.1,right] {\tiny 0} (one)
    (x4r) to node[pos=0.05,left] {\tiny 0} (zero)
    (x4r) to node[right] {\tiny 1} (one);
    \end{tikzpicture}
    
    \caption{Decision diagram representing the Boolean function $f=\bar{x}_1\bar{x}_2\bar{x}_3x_4+\bar{x}_1x_2x_3\bar{x}_4+x_1\bar{x}_2x_3\bar{x}_4+x_1x_2\bar{x}_3x_4$}
    \label{fig:dd}
\end{figure}
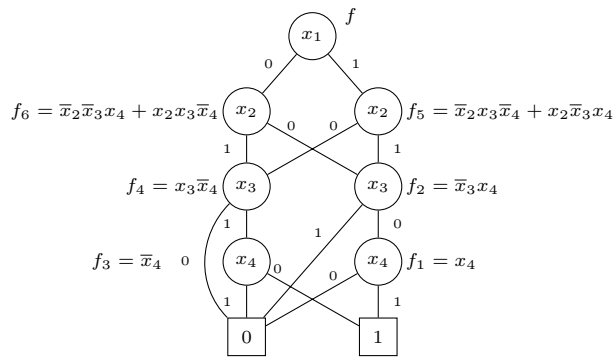

\begin{example}
Consider a Boolean function~$f:\{0,1\}^4 \to \{0,1\} =\bar{x}_1\bar{x}_2\bar{x}_3x_4+\bar{x}_1x_2x_3\bar{x}_4+x_1\bar{x}_2x_3\bar{x}_4+x_1x_2\bar{x}_3x_4$.
A straightforward \emph{complete} representation of this function would require the representation of a total of $2^4$ input-output mappings, e.g., ~in terms of a truth table.
Encoding the function as decision diagram results in a graph with only 9 nodes as illustrated in \autoref{fig:dd}.
Here, the overall function~$f$ is first decomposed with respect to variable~$x_1$ into two sub-functions~$f_6$ (assuming $x_1=0$) and~$f_5$ (assuming $x_1=1$). This is recursively continued for all remaining variables until only terminals~0 and~1 result. Whenever this decomposition yields equivalent (and, hence, redundant) sub-functions (as it is the case, e.g., for $f_4$), the sub-function is represented by a single shared node only---providing a more compact representation.
\end{example}

\begin{rem}
Albeit decision diagrams are \emph{directed} graphs, edges in illustrations do not include arrow tips since, by convention, the direction is fixed---most commonly from the top strictly to the bottom.
\end{rem}

\subsection{Proposed Decision Diagram}

In this work, we propose a type of decision diagrams aiming for 
a compact representation of a probability distribution over a given Hamiltonian.
To this end, we first start by providing the definition of the proposed type: %

\begin{defn}\label{defn:decision-diagram}
    The decision diagram we propose is a rooted directed acyclic multi-graph $G = (V, E)$ such that all maximal directed paths consist of precisely $n$ edges. Each edge $e\in E$ is equipped with two pieces of data: A traceless Pauli operator $B(e)\in\Pcal$ and a weight $w(e)\in(0,1]$ such that
    \begin{enumerate}
        \item for each vertex $v \in V$, there is at most one out-going edge for each traceless Pauli operator, and
        \item for each vertex $v\in V$ and outgoing edges $e\in\mathrm{out}(v)$, the weights are probabilistic, i.e., $\sum_{e\in\mathrm{out}(v)} w(e) = 1$ (therefore, each vertex except for the terminal has at least one out-going edge).
    \end{enumerate}
\end{defn}

Having this structure, the edge weights in the decision diagram provide the probabilities by which a random walk should follow each edge.
Multiplying the edge weight along a path gives the probability of encountering this path in a random walk.
Intuitively, the probability of a path corresponds to the weight of the Pauli operators it covers.
More precisely, the sum of the absolute values of the coefficients $| \alpha_P |$ with \(P\) covered by the path is used as relative probability and is encoded in the decision diagram.
An example illustrates the idea.

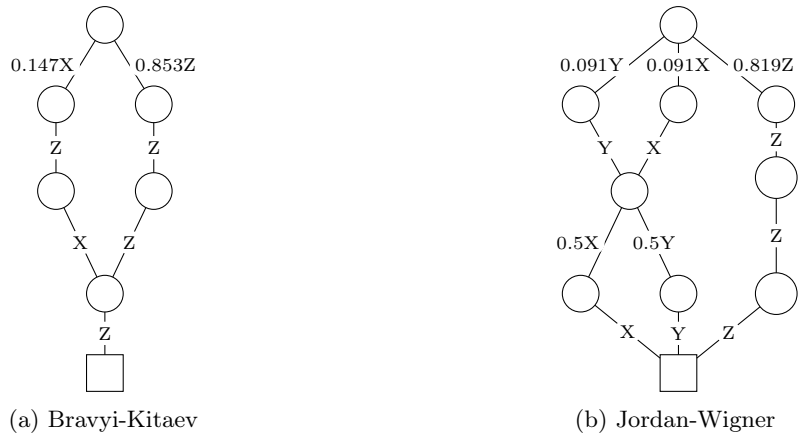
\begin{figure}
    \centering
    \begin{subfigure}{0.4\linewidth}
        \centering    
        \begin{tikzpicture}[rounded corners] 
            \graph [no placement,
                edge quotes={inner sep=2pt, font=\scriptsize, fill=white},
                nodes={circle, draw, inner sep=3pt,outer sep=0, minimum size=1.5em, text opacity=0, font=\tiny}
            ]
            {
                0 --["0.147X" left] 1[at=({-2em, -3.25em})],
                0 --["0.853Z" right] 4[at=({2em, -3.25em})],
                1 --["Z"] 2[at=({-2em, -6.8em})],
                4 --["Z"] 5[at=({2em, -6.8em})],
                2 --["X"] 3[at=({0em, -11em})],
                5 --["Z"] 3,
                3 --["Z"] t[rectangle, sharp corners, at=({0em, -14.25em})];
            };
        \end{tikzpicture}
        \caption{Bravyi-Kitaev}
        \label{fig:h2-4-bk}
    \end{subfigure}
    \quad
    \begin{subfigure}{0.4\linewidth}
        \centering
        \begin{tikzpicture}[rounded corners] 
            \graph [no placement,
                edge quotes={inner sep=2pt, font=\scriptsize, fill=white},
                nodes={circle, draw, inner sep=3pt,outer sep=0, minimum size=1.5em, text opacity=0, font=\tiny}
            ]
            {
                0 --["0.091Y" left] 1[at=({-4em, -3.25em})],
                0 --["0.091X"] 5[at=({0em, -3.25em})],
                0 --["0.819Z" right] 9[at=({4em, -3.25em})],
                1 --["Y"] 2[at=({-2em, -6.8em})],
                2 --["0.5X" left] 3[at=({-4em, -11em})],
                2 --["0.5Y"] 4[at=({0em, -11em})],
                3 --["X"] t[rectangle, sharp corners, at=({0em, -14.25em})],
                4 --["Y"] t,
                5 --["X"] 2,
                9 --["Z"] 10[at=({4em, -6.25em})],
                10 --["Z"] 11[at=({4em, -11em})],
                11 --["Z"] t;
            };
        \end{tikzpicture}
        \caption{Jordan-Wigner}
        \label{fig:h2-4-jw}
    \end{subfigure}
    \caption{Decision diagrams for H$_2$ with 4 qubits}
\end{figure}

\begin{example}
    Consider the Hamiltonian \(H\) for the hydrogen molecule H\({}_2\) with 4 qubits and Bravyi-Kitaev encoding, namely:
    \begin{align*}\scriptsize
        H = &-0.811\pt{IIII}+0.120\pt{IZII}-0.045\pt{XZXI}+0.045\pt{XIXZ}+0.045\pt{XIXI}-0.045\pt{XZXZ}\\
            &+0.120\pt{IZIZ}+0.172\pt{ZIII}-0.225\pt{IZZZ}-0.228\pt{ZZII}+0.172\pt{IIZI}+0.168\pt{ZIZI}\\
            &+0.166\pt{ZZZZ}+0.166\pt{ZZZI}+0.174\pt{ZIZZ}.
    \end{align*}
    The full-weight terms \(\pt{XZXZ}\) or \(\pt{ZZZZ}\) cover every Pauli term in the Hamiltonian, hence, the corresponding decision diagram encoding the probability distribution only has to include these two terms.
    
    \autoref{fig:h2-4-bk} illustrates a compatible decision diagram that has only two maximal paths \(\pt{XZXZ}\) and \(\pt{ZZZZ}\).
    The edges are labeled with a probability (edge weights of \(1\) are omitted for the sake of readability). 
    Using the decision diagram, one can obtain the respectively desired probabilities by traversing the decision diagram starting at the top-most node and following the edges strictly downwards until the \emph{terminal vertex} (depicted as rectangle) is reached.
    Generating measurements from this decision diagram will result in \(\pt{XZXZ}\) with a probability of 0.147 or \(\pt{ZZZZ}\) with 0.853.
    
    In \autoref{fig:h2-4-bk}, the encoded terms end with the \(Z\) operator regardless of previous choices, hence the last decision is presented by a single vertex and edge.
    This may seem like a small gain, but generally larger instances have more potential for sharing. 
    
    In a similar fashion, a decision diagram representing the Jordan-Wigner encoding can be generated---yielding the structure as shown in \autoref{fig:h2-4-jw}.
\end{example}

\subsection{Sampling Using the Proposed Decision Diagrams}

The definition in the previous subsection does not make explicit reference to the prescribed Hamiltonian. In order to use one such instance of these decision diagrams we augment the definition with the following

\begin{defn}\label{defn:compatibility}
    The decision diagram is called \emph{compatible} with the Hamiltonian $H$ if all Pauli observables $P$ with $\alpha_P\neq0$ are able to be estimated. Precisely, if $\alpha_P\neq0$ then we require at least one directed path $(e_1,\dots,e_n)$
    such that $B\in\Lift(P)$ where $B$ is the full-weight Pauli operator $\otimes_{i\in[n]} B(e_i)$.
\end{defn}

Let us make two observations that bring decision diagrams into the probabilistic setup of the shallow measurement problem as explained in the previous section.
First, the decision diagram provides a probability distribution over full-weight Pauli operators $\beta:\Pcal^n \to \RR^+$. For $B\in\Pcal^n$ such that $B=\otimes_{i\in[n]} B(e_i)$ for some maximal directed path $(e_1,\dots,e_n)$ then we set $\beta(B) = \prod_{i\in[n]} w(e_i)$. If no such maximal directed path exists, then we set $\beta(B)=0$. Condition 2 in \autoref{defn:decision-diagram} ensures that $\sum_B \beta(B)=1$.
Indeed, using the decision diagrams proposed above, samples can be drawn by perfoming a random walk.
Starting at the root vertex, a successor vertex is randomly selected according to the weights on the out-going edges.
This process is repeated at the selected vertex until the terminal vertex is reached.
Second, if the decision diagram is compatible with $H$ then, for any quantum density $\rho\in\Dcal(\Hcal)$, the estimator $\nu$ of \autoref{alg:general} is an unbiased estimator of the energy. That is, \autoref{lem:m1} establishes $\EE(\nu)=\Tr(H \rho)$.

\addtocounter{example}{-1}
\begin{example}[continued]
     Consider again \autoref{fig:h2-4-bk}.
     Sampling from this decision diagram, one starts at the root vertex and randomly chooses the \(X\) or \(Z\) edge, according to the edge weights.
     Continuing from successor vertex of the chosen edge, the remaining decisions are fixed since each following vertex only has one out-going edge, again, resulting in either \(\pt{XZXZ}\) or \(\pt{ZZZZ}\).
\end{example}

Decision diagrams provide a more powerful way to solve the shallow measurement problem. Indeed Pauli grouping via graph-coloring from \autoref{sec:rel_work} may be seen as one instance of a decision diagram according to \autoref{defn:decision-diagram} and~\ref{defn:compatibility}. Therefore any Pauli term which is covered by multiple bases under that proposal will be estimated more often under the framework here. Also, LBCS from \autoref{sec:rel_work} is a very simplistic instance of a decision diagram. We establish the decision diagram link in \hyperref[app:dd_for_old_proposals]{Appendix~\ref*{app:dd_for_old_proposals}}. As previously commented, LBCS (which observationally is better than Pauli grouping) suffers from the lack of correlation between choices of measurement bases on each qubit. The more general decision diagram framework presented here allows such correlated choices.

The larger class of distributions therefore allows us to ultimately reduce the variance associated with our estimator.
Importantly, our proposal for building decision diagrams also leads to an efficient proposal for assigning weights locally such that an attractive distribution $\beta$ is ultimately found. We reemphasize the possibility that such an algorithm could be extended further with the techniques of derandomisation.

\subsection{Optimising the Probability Distribution on the Decision Diagram}\label{subsec:optimising-beta}

It is a computationally difficult problem to find the optimal $\beta$. This would be the \emph{minimum-variance unbiased estimator} (MVUE) over all such distributions $\beta:\Pcal^n\to\RR^+$. It would be interesting to understand \emph{how close} decision diagrams get to approximating the MVUE. Nevertheless, following the method of Lagrange multipliers used in~\cite{hadfield2020} to optimise the probability distribution of LBCS, we can derive similar iterative procedure to fine tune the probability distribution $\beta$ based on the diagonal cost function described in \hyperref[sec:choosing-beta]{Appendix~\ref*{sec:choosing-beta}}. The computational cost of the iterative updates is proportional to the size of the decision diagram. 

A direct implication of the use of probability-optimised decision diagrams is to generalise and improve previous results (Theorem~3 in~\cite{evans2019scalable}~and~Theorem~1 in~\cite{huang2021efficient}) on estimating the expectation values of a collection of Pauli operators (e.g., for partial tomography~\cite{cotler2019quantum}) thanks to the ability to compute $\zeta(P, \beta)$ efficiently from a decision diagram. For example, the error bounds of previous results provide non-trivial bounds when the size of $\supp(P)$ for all Pauli $P$ is small (or, low-weight Paulis), while those of ours can give non-trivial bounds even when some of the Paulis are of full weight. We give the details in \hyperref[app:derandomisation]{Appendix~\ref*{app:derandomisation}} because the focus of this paper is on different topics of measuring quantum Hamiltonians.

\section{Efficient Construction of the Proposed Decision Diagrams}\label{sec:constructing-DD}

The decision diagrams as introduced in the previous section promise to give suitable probability distribution.
Still, the question remains how to efficiently transform the Hamiltonian consisting of coefficients and Pauli terms into a decision diagram.
In the following, we describe the main ideas. The full implementation is available at \url{https://github.com/iic-jku/dd-quantum-measurements}.

\begin{algorithm}[tb]
	\caption{Construction of a decision diagram (DD) from Hamiltonian \(H\)}
	\label{alg:dd-construction}
	\begin{algorithmic}
        \State Take absolute values of coefficients in \(H\)
        \State Merge compatible terms to get reduced positive Pauli list \(\Rcal(H)\) \Comment{Preprocessing}
        \For{Each term and coefficient in \(\Rcal(H)\)} \Comment{Initialisation of DD}
            \State Take existing path covering the longest prefix of term
            \State Create new edges for remaining Pauli operators up to the last
            \State Create edge to terminal with the last Pauli op and coefficient as edge weight
        \EndFor
        \For{Vertex in decision diagram in breadth-first order from terminal}\Comment{Normalisation of DD}
            \State Calculate sum of weights on out-going edges
            \State Divide weights on out-going edges by sum and multiply sum to in-coming edge weights
        \EndFor
        \For{Vertex in decision diagram in breadth-first order from terminal}\Comment{Merge equivalent vertices in DD}
            \State Calculate hash of vertex and if equivalent vertex exists, merge both
        \EndFor
        \State Remove identities in DD
        \State \hspace{\algorithmicindent} Replace \enquote{lonely} identity edges with virtual edges
        \State \hspace{\algorithmicindent} Remove identity edges where other edge with same source and target exists
        \State \hspace{\algorithmicindent} Merge targets of identity edges with target vertices of other edge
        \For{Vertex in decision diagram in breadth-first order from terminal}\Comment{Merge equivalent vertices in DD}
            \State Calculate hash of vertex and if equivalent vertex exists, merge both
        \EndFor
	    \State \Return decision diagram
	\end{algorithmic}
\end{algorithm}

In a high-level view, the construction of the proposed decision diagram consists of multiple steps as shown in \autoref{alg:dd-construction}.
The first stage is a preprocessing step of the Hamiltonian to reduce the number of identity-terms.
The second stage is the initialisation and refinement of the decision diagram.
This second stage has several steps. Only after these steps have been performed are we guaranteed that the decision diagram conforms to both \autoref{defn:decision-diagram} and \autoref{defn:compatibility}. These steps include normalising the information present in the prepocessed Hamiltonian in order to maximise sharing and also removing identity-edges through \emph{merging}.
The following paragraphs explain the individual steps in more detail.

\begin{figure}
    \centering
    \begin{subfigure}{0.4\linewidth}
        \centering
        \begin{tikzpicture}[rounded corners] 
            \graph [no placement,
                edge quotes={inner sep=2pt, font=\scriptsize, fill=white},
                nodes={circle, draw, inner sep=3pt,outer sep=0, minimum size=1.5em, text opacity=0, font=\tiny}
            ]
            {
                 0 --["Y"]  1[at=({-6em, -3.25em})],
                 0 --["X"]  5[at=({0em, -3.25em})],
                 0 --["Z"]  9[at=({6em, -3.25em})],
                 1 --["Y"]  2[at=({-6em, -6.8em})],
                 2 --["X"]  3[at=({-10em, -11em})],
                 2 --["Y"]  4[at=({-6em, -11em})],
                 3 --[bend right, "0.045X" left] t[rectangle, sharp corners, at=({-2em, -14.25em})],
                 4 --["0.045Y" left] t,
                 5 --["X"]  6[at=({0em, -6.8em})],
                 6 --["X"]  7[at=({-2em, -11em})],
                 6 --["Y"]  8[at=({2em, -11em})],
                 7 --["0.045X"]  t,
                 8 --["0.045Y" right]  t,
                 9 --["Z"] 10[at=({6em, -6.8em})],
                10 --["Z"] 11[at=({6em, -11em})],
                11 --[bend left, "1.714Z"]  t;
            };
        \end{tikzpicture}
        \caption{Initial DD}
        \label{fig:init-dd}
    \end{subfigure}
    \quad
    \begin{subfigure}{0.4\linewidth}
        \centering
        \begin{tikzpicture}[rounded corners] 
            \graph [no placement,
                edge quotes={inner sep=2pt, font=\scriptsize, fill=white},
                nodes={circle, draw, inner sep=3pt,outer sep=0, minimum size=1.5em, text opacity=0, font=\tiny}
            ]
            {
                 0 --["0.048Y"]  1[at=({-6em, -3.25em})],
                 0 --["0.048X"]  5[at=({0em, -3.25em})],
                 0 --["0.904Z" xshift=6pt]  9[at=({6em, -3.25em})],
                 1 --["Y"]  2[thick, at=({-6em, -6.8em})],
                 2 --["0.5X"]  3[at=({-10em, -11em})],
                 2 --["0.5Y"]  4[at=({-6em, -11em})],
                 3 --["X"]  t[rectangle, sharp corners, at=({-2em, -14.25em})],
                 4 --["Y"]  t,
                 5 --["X"]  6[thick, at=({0em, -6.8em})],
                 6 --["0.5X"]  7[at=({-2em, -11em})],
                 6 --["0.5Y"]  8[at=({2em, -11em})],
                 7 --["X"]  t,
                 8 --["Y"]  t,
                 9 --["Z"] 10[at=({6em, -6.8em})],
                10 --["Z"] 11[at=({6em, -11em})],
                11 --["Z"]  t;
            };
            \draw[shorten <=3pt, shorten >=3pt, <->] (6) edge node[below, font=\tiny]{equivalent} (2);
        \end{tikzpicture}
        \caption{Normalised DD}
        \label{fig:dd-normalized}
    \end{subfigure}
    \caption{First steps of decision diagram construction from reduced positive Hamiltonian of H\(_2\) JW (4 qubits)}
    \label{fig:dd-steps-i}
\end{figure}

\emph{Preprocessing:}
Consider the Hamiltonian as presented in \autoref{eqn:hamiltonian}. 
Immediately, we remove the term $\alpha_{I^n} I^n$ as this term does not need to be estimated using the quantum processor. 
We also map coefficients to their absolute values (\(\alpha_P \mapsto |\alpha_P|\)) giving what we will call the \emph{positive Pauli list}.
Then, compatible Pauli terms in this positive Pauli list are merged to reduce the number of paths in the initial decision diagram.
More precisely, for each of Pauli terms the number of compatible terms is determined. The Pauli term \(P_\mathrm{high}\) with the highest number of compatible terms \(n_\mathrm{comp}\) is subsequently merged into these compatible terms and the fraction \(\frac{\alpha_{P_\mathrm{high}}}{n_\mathrm{comp}}\) is added to the coefficient of each compatible term.
This merging procedure is repeated until no further merging is possible.
This preprocessing provides what we shall refer to as the \emph{reduced positive Pauli list} and shall denote it by $\Rcal(H)$.

\emph{Initialisation of DD:}
From the reduced positive Pauli list $\Rcal(H)$, an initial decision diagram whose maximal paths are all of length $n$ is constructed. 
Each term in $\Rcal(H)$ is associated with a unique maximal path and the coefficients of $\Rcal(H)$ are assigned to the final edges in the respective path, i.e.,~to the edge pointing to the terminal vertex.

\emph{Normalisation of DD:}
Afterwards the edge weights are normalised such that the sum of weight of out-going edges equals 1.
This decision diagram at this stage has sharing for common prefixes (but not suffixes) and at this point may include edges with the identity operator.

\begin{example}
    Consider the 4 qubit Hamiltonian for the hydrogen molecule in Jordan-Wigner encoding:
    \begin{align*}
        H = &-0.810\pt{IIII}+0.045\pt{YYXX}+0.045\pt{YYYY}+0.045\pt{XXXX}\\
            &+0.045\pt{XXYY}+0.172\pt{ZIII}-0.225\pt{IZII}+0.172\pt{IIZI}\\
            &-0.225\pt{IIIZ}+0.120\pt{ZZII}+0.168\pt{ZIZI}+0.166\pt{ZIIZ}\\
            &+0.166\pt{IZZI}+0.174\pt{IZIZ}+0.120\pt{IIZZ}
    \end{align*}
    The reduced positive Pauli list from \(H\) is
    \begin{align*}
        \Rcal(H) = 0.045\pt{YYXX}+0.045\pt{YYYY}+0.045\pt{XXXX}+0.045\pt{XXYY}+1.714\pt{ZZZZ}
    \end{align*}
    \autoref{fig:init-dd} illustrates the initial decision diagram created from \(\Rcal(H)\) and \autoref{fig:dd-normalized} illustrates the same decision diagram with normalised edge weights.
\end{example}

Before considering removing potentially remaining identity edges in the decision diagram, functionally equivalent vertices are merged.
Two vertices are equivalent if they have the same successors considering the Pauli operator and respective weight.
Given a suitable hash function, finding equivalent nodes is linear in the number of nodes~\cite{DBLP:conf/iccad/ZulehnerHW19}.

\addtocounter{example}{-1}
\begin{example}[continued]
    Consider again \autoref{fig:dd-normalized}.
    The two vertices highlighted by a bold border are equivalent and, thus, are combined to exploit the redundancy.
    This results in the decision diagram as already shown before in \autoref{fig:h2-4-jw}.
\end{example}

\begin{figure}
    \centering
    \begin{subfigure}{0.4\linewidth}
        \centering
        \begin{tikzpicture}[rounded corners] 
            \begin{scope}[local bounding box=graph1]
            \graph [no placement,
                edge quotes={auto, inner sep=1pt, font=\scriptsize},
                nodes={circle, draw, inner sep=3pt,outer sep=0, minimum size=1.5em, text opacity=0, font=\tiny}
            ] 
            {
                 0 --[bend right, "0.5I" left]  1[at=({0em, -3.5em})],
                 0 --[bend left, "0.5X"]  1;
            };
            \end{scope}

            \begin{scope}[local bounding box=graph2, xshift=2cm]
            \graph [no placement,
                edge quotes={inner sep=2pt, font=\scriptsize, fill=white},
                nodes={circle, draw, inner sep=3pt,outer sep=0, minimum size=1.5em, text opacity=0, font=\tiny}
            ]
            {
                 0 --["X"]  1[at=({0em, -3.5em})];
            };
            \end{scope}
            \node at ($(graph1)!0.6!(graph2)$) {\(\leadsto\)};
        \end{tikzpicture}
        \caption{Combining identities}
        \label{fig:combine-identities}
    \end{subfigure}
    \quad
    \begin{subfigure}{0.4\linewidth}
        \centering
        \begin{tikzpicture}[rounded corners] 
            \begin{scope}[local bounding box=graph1]
            \graph [no placement,
                edge quotes={inner sep=2pt, font=\scriptsize, fill=white},
                nodes={circle, draw, inner sep=3pt,outer sep=0, minimum size=1.5em, text opacity=0, font=\tiny}
            ] 
            {
                 0 --["I"]  1[at=({0em, -3.5em})],
            };
            \end{scope}

            \begin{scope}[local bounding box=graph2, xshift=2.5cm]
            \graph [no placement,
                edge quotes={inner sep=2pt, font=\scriptsize, fill=white},
                nodes={circle, draw, inner sep=3pt,outer sep=0, minimum size=1.5em, text opacity=0, font=\tiny}
            ]
            {
                 0 --[bend right=75, "\(\frac{1}{3}\dot{X}\)" left]  1[at=({0em, -3.5em})];
                 0 --["\(\frac{1}{3}\dot{Y}\)"]  1;
                 0 --[bend left=75, "\(\frac{1}{3}\dot{Z}\)" right]  1;
            };
            \end{scope}
            \node at ($(graph1)!0.4!(graph2)$) {\(\leadsto\)};
        \end{tikzpicture}
        \caption{Split into \emph{virtual} edges}
        \label{fig:split-identities}
    \end{subfigure}
    \caption{Removing identity edges from the decision diagram}
    \label{fig:dd-steps-ii}
\end{figure}
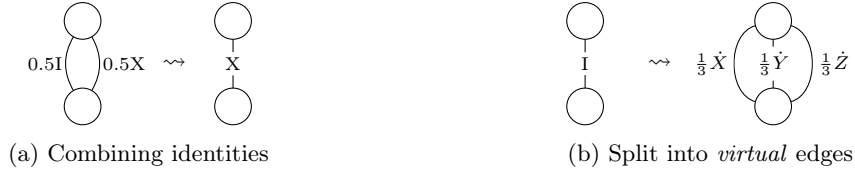

After the previous steps, the decision diagram may still have identity edges, which have to be removed to get a proper probability distribution over the Hamiltonian.
The potentially remaining identity edges are removed in three steps.

The first two steps are local operations. Given two fixed vertices $u,v \in V$ the following checks are performed:
\begin{enumerate}
    \item For fixed $u,v\in V$, if there is an edge $u\xrightarrow[]{I}v$ and any $u\xrightarrow[]{\{X,Y,Z\}}v$: Remove $u\xrightarrow[]{I}v$ and add the weight to the remaining edge from $u\xrightarrow[]{\{X,Y,Z\}}v$ with the smallest weight. \autoref{fig:combine-identities} illustrates this step in an example with one further edge.
    \item Again, for fixed $u, v \in V$, if there is only one edge $u\xrightarrow{I}v$ (referred to as \enquote{lonely} in \autoref{alg:dd-construction}): Split $u\xrightarrow{I}v$ into \emph{virtual edges} (denoted by a dot above the operator) $u\xrightarrow{\dot{X}}v$, $u\xrightarrow{\dot{Y}}v$, and $u\xrightarrow{\dot{Z}}v$ with weights $\frac{1}{3}$ each. \autoref{fig:split-identities} illustrates this step.
\end{enumerate}

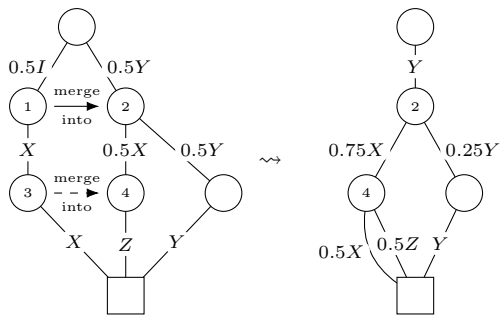
\begin{figure}
    \centering
    \begin{tikzpicture}[rounded corners] 
        \begin{scope}[local bounding box=graph1]
        \graph [no placement,
            edge quotes={inner sep=2pt, font=\scriptsize, fill=white},
            nodes={circle, draw, inner sep=3pt,outer sep=0, minimum size=1.5em, text opacity=1, font=\tiny}
        ] 
        {
             0[text opacity=0] --["\(0.5I\)" left]  1[at=({-2em, -3.25em})],
             0 --["\(0.5Y\)" right]  2[at=({2em, -3.25em})],
             1 --["\(X\)"]  3[at=({-2em, -6.8em})],
             2 --["\(0.5X\)"]  4[at=({2em, -6.8em})],
             2 --["\(0.5Y\)" right]  5[text opacity=0, at=({6em, -6.8em})],
             3 --["\(X\)"]  t[text opacity=0, rectangle, sharp corners, at=({2em, -11em})],
             4 --["\(Z\)"]  t,
             5 --["\(Y\)"]  t;
        };
        \draw[shorten <=3pt, shorten >=3pt, -Latex] (1) edge node[above, font=\tiny]{merge} node[below, font=\tiny]{into} (2);
        \draw[dashed, shorten <=3pt, shorten >=3pt, -Latex] (3) edge node[above, font=\tiny]{merge} node[below, font=\tiny]{into} (4);
        \end{scope}

        \begin{scope}[local bounding box=graph2, xshift=4.5cm]
        \graph [no placement,
            edge quotes={inner sep=2pt, font=\scriptsize, fill=white},
            nodes={circle, draw, inner sep=3pt,outer sep=0, minimum size=1.5em, text opacity=1, font=\tiny}
        ]
        {
             0[text opacity=0] --["\(Y\)"]  2[at={(0,-3.25em)}],
             2 --["\(0.75X\)" left]  4[at={(-2em,-6.8em)}],
             2 --["\(0.25Y\)" right]  5[at={(2em,-6.8em)}, text opacity=0],
             5 --["\(Y\)"]  t[at={(0,-11em)}, text opacity=0, rectangle, sharp corners],
             4 --["\(0.5Z\)" xshift=3pt]  t,
             4 --[bend right, "\(0.5X\)" left]  t,
        };
        \end{scope}
        \node at ($(graph1.east)!0.4!(graph2.west)$) {\(\leadsto\)};
    \end{tikzpicture}
    \caption{Merging two vertices}
    \label{fig:merging-vertices}
\end{figure}

The remaining identity edges cannot be removed by only considering individual pairs of nodes, but require a more global approach.
Recall that at this point there are no two vertices with only an identity edge between them. 
So for $u\xrightarrow[]{I}v$ and any $u\xrightarrow[]{\{X,Y,Z\}}v'$ we merge $v$ into $v'$ and adjust the weights accordingly.
More precisely, the merging is handled by checking the following list for each out-going edge of \(v\):
\begin{enumerate}
    \item If the target vertex \(v'\) does not have an out-edge with the same Pauli operator as the currently considered out-edge of \(v\), add this edge to \(v'\).
    \item If the currently considered out-edge of \(v\) and the out-edge of \(v'\) with the same operator point to same vertex (which may be the terminal vertex), the weights stay the same.
    \item Otherwise the merging process has to recurse to merge the successors of \(v\) and \(v'\) with corresponding Pauli operators.
\end{enumerate}
During the merging of two edges, the resulting edge is only \emph{virtual} if both previous edges were virtual.
Performing the merging process from the terminal vertex upwards ensures that the recursive applications never encounter an identity edge.
At this point there may be superfluous virtual edges left, which are removed to reduce the number of paths.
After the merging process is completed, the decision diagram is renormalised to ensure the sum of weights of out-going edges on a node equals 1.
An example illustrates the idea.
\begin{example}
    Consider the left-hand side of \autoref{fig:merging-vertices}.
    This decision diagram has a top vertex with two out-going edges with the operators \(I\) and \(Y\).
    To remove the identity edge, the target vertex of the identity edge (1) is merged into the target vertex of the Pauli-Y edge (2) (indicated by a solid an arrow labeled \enquote{merge into}).
    Since both (1) and (2) have an out-going Pauli-X edge to different target vertices, theses targets (3) and (4) have to merged as well in a recursive fashion (indicated by a dashed arrow).
\end{example}

The following section provides the experimental evaluation on decision diagrams constructed as described above.

\section{Numerical Experiments}
\label{sec:experimental-evalutation}

In this section, we give numerical experiments demonstrating the efficacy of decision diagrams (DDs) to solve the shallow-circuit measurement problem.
In particular, the experiments are to confirm that DDs allow to encode probability distributions $\beta$ from a wider class of probabilities than
with known approaches, such as, the Pauli grouping and LBCS. Following the experimental setting of~\cite{hadfield2020}, we consider five molecular Hamiltonians that range 
from 4 to 14 spin orbitals. Excepting the 8-qubit H$_2$ (hydrogen) that uses a 6—31G basis, all Hamiltonians are represented in a minimal STO-3G basis. 
Each of the molecular Hamiltonians is turned into a qubit Hamiltonian using three encodings: Jordan-Wigner (JW), Parity, and Bravyi-Kitaev (BK)~\cite{bravyi2017tapering}. 
The resulting qubit Hamiltonians are those requiring 4 to 14 qubits that are defined in the molecular basis, and therefore their Hartree-Fock states are those of computational basis. 
We compare the variances of the measurements obtained from decision diagrams against those from the Pauli grouping and LBCS as shown in \autoref{tab:encodings}.

In the table we list the variances of estimators of two established approaches (LDF, LBCS) and four proposed DD-based approaches:
\begin{itemize}
    \item LDF Pauli grouping~\cite{hadfield2020} 
    \item Locally-biased classical shadows~\cite{hadfield2020}
    \item Decision diagrams constructed as described in \autoref{sec:DD estimators}.
    \item Decision diagrams with the same construction but up to ten optimisation passes as described in \autoref{subsec:optimising-beta} (denoted as \textit{DD+opt10})
    \item Decision diagrams with paths constructed from the LDF Pauli grouping (denoted as  \textit{DD+LDF})
    \item The DD+LDF with ten optimisation passes as described in \autoref{subsec:optimising-beta} (denoted as \textit{DD+LDF+opt10})
\end{itemize}

As shown in \autoref{tab:encodings}, the variances of the DD-based probability distributions are lower than those of LBCS for H$_2$, LiH, and BeH$_2$, but show deterioration for larger molecules such as H$_2$O. 
For the smallest considered molecule (4-qubit H$_2$), DDs provide a slightly lower variance for the JW encoding compared to LDF grouping but consistently beat LBCS.
Also, we can confirm the efficacy of the DDs from the variances of DD+LDF which are always better than those of Pauli grouping for all considered molecules, except 4-qubit H$_2$ in the Parity and BK encodings. 

The variances for DD-based probability distributions can be further reduced by applying the optimisation procedure from \autoref{subsec:optimising-beta}, as can be seen from the column of \textit{DD+opt10} and \textit{DD+LDF+opt10}: 
Even with limited steps of optimising the probability distribution, the variances can be reduced below those of LBCS for almost all considered molecules and encodings (except H$_2$O in the JW encoding). %
From the column of \textit{DD+LDF+op10} we can observe that although the variances are improved by tuning the probability distribution, the variances are mostly worse than those of \textit{DD+opt10}, thus demonstrating the effectiveness of our proposed construction of the DD against the LDF-based one. 
Apart from the attained variances, decision diagrams are commonly characterized by their number of vertices, number of edges and number of paths.
\autoref{tab:dd-metrics} provides these details for the respective molecules, encoding, and (optimised) approaches.
In particular, \textit{DD} generates more compact decision diagrams (i.e.,~lower number of vertices) compared to \textit{DD+LDF}, but \textit{DD} generates more or at least the same number of paths (excepting for LiH at BK encoding). This hints at the importance of designing compact DDs for shallow measurements.

\begin{table}[tbp]
    \centering
    \caption{Variance for different estimators computed on the ground states}
    \label{tab:encodings}
    \begin{tabular}{ll@{\hspace{1em}}S[table-format=4.3]S[table-format=4.3]@{\hspace{1em}}S[table-format=4.3]S[table-format=4.3]S[table-format=4.3]S[table-format=4.3]}
    	\toprule
    	         &          &                \multicolumn{6}{c}{Variance}                 \\
    	\cmidrule{3-8}
    	Molecule & Encoding & {LDF grouping} & {LBCS} & {DDs}   & {DD+opt10} & {DD+LDF}  & {DD+LDF+opt10} \\ \midrule
    	H$_2$ (4 qubits)    & JW       & 0.402          & 1.860  & 0.361   & 0.398      & 0.361     & 0.398 \\
    	                    & Parity   & 0.193          & 0.541  & 0.307   & 0.300      & 0.292     & 0.285 \\
    	                    & BK       & 0.193          & 0.541  & 0.307    & 0.300      & 0.292     & 0.285 \\[6pt]
    	H$_2$ (8 qubits)    & JW       & 22.3           & 17.7   &  8.7    & 6.2        & 13.2      & 7.6 \\
    	                    & Parity   & 38.0           & 18.9   & 10.2    & 8.5        & 13.6      & 8.9 \\
    	                    & BK       & 38.4           & 19.5   &  7.5    & 6.4        & 16.2      & 9.1 \\[6pt]
    	LiH (12 qubits)     & JW       & 54.2           & 14.8   & 13.5   & 8.5     & 33.1   & 16.3 \\
    	                    & Parity   & 85.8           & 26.5   & 24.2   & 12.5    & 53.8   & 27.2 \\
    	                    & BK       & 75.5           & 68.0   & 31.0   & 14.2    & 72.0   & 37.4 \\[6pt]
    	BeH$_2$ (14 qubits) & JW       & 135            & 67.6   & 51.8   & 32.8    & 68.4   & 35.1 \\
    	                    & Parity   & 239            & 130    & 72.5   & 37.2    & 96.0   & 55.8 \\
    	                    & BK       & 197            & 238    & 200.1  & 64.2    & 147.9  & 73.2 \\[6pt]
    	H$_2$O (14 qubits)  & JW       & 1040           & 258    & 616.7  & 294.4    & 829.4  & 336.3 \\
    	                    & Parity   & 2670           & 429    & 915.6  & 425.3    & 660.9  & 368.0 \\
    	                    & BK       & 2090           & 1360   & 1084.8 & 527.0    & 1403.4 & 628.1 \\
        \bottomrule
    \end{tabular}
\end{table}

\begin{table}[tbp]
    \centering
    \caption{Metrics on the generated decision diagrams}    
    \label{tab:dd-metrics}
    \begin{tabular}{lll@{\hspace{2em}}S[table-format=4.0]S[table-format=4.0]S[table-format=4.0]}
    	\toprule
    	Molecule         & Encoding & Approach  &    {|Vertices|} &    {|Edges|} & {|Paths|} \\ \midrule
    	H$_2$ (4 qubits)    & JW       & DD     &   10 &   12 &     5 \\
    	                 &        & DD+LDF &   10 &   12 &     5 \\[3pt]
    	                 & Parity   & DD     &    7 &    7 &     2 \\
    	                 &    & DD+LDF &    7 &    7 &     2 \\[3pt]
    	                 & BK       & DD     &    7 &    7 &     2 \\
    	                 &        & DD+LDF &    7 &    7 &     2 \\[6pt]
    	H$_2$ (8 qubits)    & JW       & DD     &   77 &  133 &    77 \\
    	                 &        & DD+LDF &   84 &  130 &    61 \\[3pt]
    	                 & Parity   & DD     &   66 &  119 &    55 \\
    	                 &    & DD+LDF &   54 &   85 &    34 \\[3pt]
    	                 & BK       & DD     &   60 &  101 &    46 \\
    	                 &        & DD+LDF &   56 &   88 &    34 \\[6pt]
    	LiH (12 qubits)  & JW       & DD     &  191 &  326 &   273 \\
    	                 &        & DD+LDF &  321 &  459 &   151 \\[3pt]
    	                 & Parity   & DD     &  289 &  523 &   315 \\
    	                 &    & DD+LDF &  381 &  550 &   182 \\[3pt]
    	                 & BK       & DD     &  413 &  666 &   673 \\
    	                 &        & DD+LDF &  518 &  749 &  1012 \\[6pt]
    	BeH$_2$ (14 qubits) & JW       & DD     &  537 &  761 &   289 \\
    	                 &        & DD+LDF &  552 &  693 &   147 \\[3pt]
    	                 & Parity   & DD     &  570 &  872 &  1508 \\
    	                 &    & DD+LDF &  636 &  816 &   192 \\[3pt]
    	                 & BK       & DD     &  494 &  746 &  2192 \\
    	                 &        & DD+LDF &  858 & 1051 &   222 \\[6pt]
    	H$_2$O (14 qubits)  & JW       & DD     &  556 &  894 &  1013 \\
    	                 &        & DD+LDF &  759 &  983 &   234 \\[3pt]
    	                 & Parity   & DD     &  759 & 1314 &  5711 \\
    	                 &    & DD+LDF &  762 & 1029 &   290 \\[3pt]
    	                 & BK       & DD     &  800 & 1271 &  3914 \\
    	                 &        & DD+LDF & 1083 & 1395 &   618 \\ \bottomrule
    \end{tabular}
    \vspace{-1em}
\end{table}

\section{Conclusion}
\label{sec:conclusion}

We have introduced a new estimator for measuring quantum operators defined as linear combination of tensor products of single-qubit Pauli operators. The estimator is defined within a probabilistic measurement framework, where single-qubit measurement bases are drawn from probability distributions obtained using decision diagrams. The decision diagrams used to sample from measurement bases are constructed from target quantum operators, typically Hamiltonians, by associating paths in the diagrams with Pauli operators present in the Hamiltonians. The diagrams can then be simplified by removing paths with identities operators, and merging equivalent sub-paths.  

We have shown that representing probability distributions with decision diagrams generalises previous classical-shadow randomises approaches to the measurement problem, namely the uniform~\cite{Huang2020PredictingMeasurements} and the locally-biased one~\cite{hadfield2020}.
This generalisation comes with additional degrees of freedom that characterize each diagram, and introduce correlations between measurement bases for each qubit. We presented different strategies to optimise these additional degrees of freedom, and have shown numerically that they can outperform locally-biased approaches as well as Pauli grouping strategies, on selected molecular Hamiltonian models. 
We foresee that future refined approaches in the construction and optimisation of the diagrams could further improve on the improvements in estimation precision reported here, especially considering problem-specific decision diagram construction methods.

\subsection*{Acknowledgements}

SH and RW have received funding from the European Research Council (ERC) under the European Union’s Horizon~2020 research and innovation programme (grant agreement No.~101001318), the LIT~Secure and Correct Systems Lab funded by the State of Upper Austria, as well as the BMK, BMDW, and the State of Upper Austria in the context of the COMET program (managed by the FFG). RR would like to thank Sergey Bravyi and Shigeru Yamashita for technical discussion.  

\bibliographystyle{alpha}
\bibliography{references}

\appendix

\section{Calculation of Variance}\label{app:variance}
In this appendix we provide a detailed calculation of the variance of the estimator for $\Tr(H\rho)$ in \autoref{alg:general}. We shall set the number of preparations of the state $\rho$ to be $S=1$ in order to simplify the exposition. (We may thus omit the superscript ${(s)}$.)

For ease of readership we reintroduce some of the notation. Set $\Pcal:=\{X,Y,Z\}$ and consider a full-weight Pauli operator $B\in\Pcal^n$. Let $\mu(B,i)\in\{\pm 1\}$ denote the eigenvalue measurement when qubit $i$ is measured in the $B_i$ basis. For a subset $A\subseteq[n]$ declare
$\mu(B,A) := \prod_{i\in A} \mu(B,i)$.
For a Pauli operator $P\in\{I,X,Y,Z\}^n$, we set
\begin{equation}
    \Lift(P) :=
    \left\{
        B \in \Pcal^n
        \,|\,
        B_i = P_i \textrm{ for every }i\in\supp(P)
    \right\}.
\end{equation}
For a probability distribution on full-weight Pauli operators $\beta:\Pcal^n\to\RR^+$ we write
\begin{equation}
    \zeta(P,\beta)
    :=
    \sum_{B\in\Lift(P)}\beta(B).
\end{equation}\label{eq:zeta_P_beta}
A distribution is compatible with $H$ if $\zeta(P,\beta)>0$ whenever $\alpha_P\neq0$. From now on, we shall assume that any distribution that we consider is compatible with the Hamiltonian. (This will avoid division-by-zero errors in our subsequent calculations.)
The indicator function $\indicator$, for a set $\Omega$, returns $\indicator_\Omega(x)=1$ if $x\in\Omega$ and $\indicator_\Omega(x)=0$ if $x\not\in\Omega$. We finish this paragraph with \autoref{alg:general} rewritten with the simplifying assumption that the number of preparations of $\rho$ is $S=1$.

\begin{algorithm}[h]
	\caption{Shallow-measurement estimation of $\Tr(H\rho)$ given $\beta$ and single preparation of $\rho$.}
	\label{alg:general_single_shot}
	\begin{algorithmic}
			\State Prepare $\rho$
			\State Select basis $B\in\Pcal^n$ from $\beta$-distribution
			\For{qubit $i\in[n]$}
				\State Measure qubit $i$ in basis $B_i$ giving $\mu(P,i) \in \{\pm 1\}$
			\EndFor
		\Return
			\[
			\nu = 
			\sum_P \alpha_P 
			\cdot 
			\frac{ \indicator_{\Lift(P)} (B) }{ \zeta(P,\beta) }
			\cdot
			\mu(B,\supp(P))
			\]
	\end{algorithmic}
\end{algorithm}

In order to calculate the variance, we start by establishing that $\nu$ is unbiased.

\begin{proof}[Proof of \autoref{lem:m1}]
Let $\EE_B$ denote the expected value over the distribution $\beta(B)$.
Let $\EE_{\mu(B)}$ denote the expected value over the measurement outcomes
for a fixed Pauli basis $B$. By definition, the expected value in \autoref{eq:moment1} is a composition
of the expected values over a Pauli basis $B$ and over the measurement outcomes $\mu(B)$, that is,
$\EE = \EE_B \EE_{\mu(B)}$. There are two necessary observations in this language given an operator $P\in\{I,X,Y,Z\}^n$.
First, the indicator function has the property that $\EE_B (\indicator_{\Lift(P)}(B))=\zeta(P,\beta)$. Second if $B\in\Lift(P)$, then $\EE_{\mu(B)} \mu(B,\supp(P)) =\Tr(P\rho)$. 
Combining these observations implies
\begin{align*}
    \EE(\nu)
    &=
    \EE_B \EE_{\mu(B)} \nu
    \\
    &=
    \sum_P 
        \alpha_P 
        \frac1{\zeta(P,\beta)} 
        \EE_B ({\indicator_{\Lift(P)}}(B))
        \EE_{\mu(B)} \mu(B,\supp(P))
    \\
    &=
    \sum_P 
        \alpha_P 
        \frac1{\zeta(P,\beta)}
        \zeta(P,\beta)
        \cdot
        \Tr(P\rho)
    \\
    &=
    \sum_P \alpha_P \Tr(P\rho).
    \qedhere
\end{align*}
\end{proof}

For Pauli operators $P,Q\in\{I,X,Y,Z\}^n$, define
\begin{equation}
    g(P,Q,\beta)
    :=
    \frac1{\zeta(P,\beta)}
    \frac1{\zeta(Q,\beta)}
    \sum_{B\in\Lift(P)\cap\Lift(Q)}
        \beta(B)
\end{equation}

\begin{rem}\label{rem:var_lbcs}
This function simplifies greatly when $\beta$ is a product distribution. Specifically, if $\beta=\prod_{i=1}^n \beta_i$ with $\beta_i:\Pcal\to\RR^+$, then $g(P,Q,\beta)$ is non-zero only when $P,Q$ agree with each-other on $A=\supp(P)\cap\supp(Q)$ and in this case $g(P,Q,\beta) = \left( \prod_{i\in A} \beta_i(P_i) \right)^{-1}$.
\end{rem}

\begin{proof}[Proof of \autoref{lem:m2}]
We use the same notation as in the preceding lemma. 
Consider $P,Q\in\{I,X,Y,Z\}^n$. As operators, we obtain the identity
\begin{equation}
    \EE_B
    \frac{ \indicator_{\Lift(P)} (B) }{ \zeta( P, \beta ) }
    \frac{ \indicator_{\Lift(Q)} (B) }{ \zeta( Q, \beta ) }
    =
    g(P,Q,\beta)
\end{equation}
and, whenever $B\in\Lift(P)\cap\Lift(Q)$,
\begin{equation}
    \EE_{\mu(B)} \mu(B,\mathrm{supp}(P)) \mu(B,\mathrm{supp}(Q))
    =
    \Tr(PQ \rho).
\end{equation}
To get the last equality, observe that $\mu(B,A)\mu(B,A')=\mu(B,A\ominus A')$
for any subsets of qubits $A,A'$, where $A\ominus A'$ is the symmetric difference of $A$ and $A'$.
The assumption that $B$ is covered by both $P$ and $Q$ implies that $\mathrm{supp}(P)\ominus \mathrm{supp}(Q)=\mathrm{supp}(PQ)$.

Combining these observations implies
\begin{align}
    \EE(\nu^2)
    &=
    \EE_B \EE_{\mu(B)} \nu^2
    \nonumber\\
    &=
    \sum_{P,Q} \alpha_P \alpha_Q 
        \EE_B 
        \frac{ \indicator_{\Lift(P)} (B) }{ \zeta( P, \beta ) }
        \frac{ \indicator_{\Lift(Q)} (B) }{ \zeta( Q, \beta ) }
            \EE_{\mu(B)} \mu(B,\supp(P))\mu(B,\supp(Q))
    \nonumber\\
    &=
    \sum_{P,Q} \alpha_P \alpha_Q 
        g(P,Q,\beta)
            \Tr(PQ \rho).\label{eqn:app-variance-cost}
    \qedhere
\end{align}
\end{proof}

\section{Data structure implementation of Algorithm}\label{app:data_structure}

Consider \autoref{alg:general}. We explicit how this algorithm would be implemented in software. This implementation shows that the calculation of $\zeta(P,\beta) = \sum_{B\in\Lift(P)}\beta(B)$ is not performed inside the algorithm. The input to the eventual function is: the Hamiltonian $H$, the state $\rho$, the distribution $\beta$, the number of preparations $S$.

For a Hamiltonian $H=\sum_P \alpha_P P$, create a dictionary \verb|Estimates| whose keys are $P\in\{I,X,Y,Z\}^n$ such that $\alpha_P\neq0$. The value of $\verb|Estimates[P]|$ is a 2-tuple $(n_P, \mu_P)$ which will be updated after each preparation-and-measurement. Here, $\mu_P$ is our current best estimate of $\Tr(P\rho)$ and $n_P$ is the current number of times that a basis has been chosen which allows us to give a non-zero estimate $\Tr(P\rho)$, that is, the number of times that a basis has been chosen from $\Lift(P)$. (We therefore initialise $n_P, \mu_P$ to be zero.)

For each shot $s\in[S]$, choose a basis $B$ from $\beta$-distribution and measure each qubit in the $B_i$ basis. Now loop over all Pauli $P$ with $\alpha_P\neq0$. For one such $P$, check if $B\in\Lift(P)$. If $B\not\in\Lift(P)$ then move to the next Pauli. If $B\in\Lift(P)$ then we will update $(n_P,\mu_P)$ according to
\begin{align}
    \mu_P &\leftarrow \frac1{n_P+1}\left( \mu(B,\supp(P)) + n_P \mu_P \right)
    \\
    n_P &\leftarrow n_P + 1
\end{align}

The function in software then returns $\sum_P \alpha_P \mu_P$.

\section{Previous Sampling Approaches in Decision Diagrams}\label{app:dd_for_old_proposals}

In this section we illustrate how the methods of Locally Biased Classical Shadows (LBCS) and Pauli grouping may be seen as instances of Decision Diagrams. We then 
show how the representation of Decision Diagrams may be used to improve 
both existing sampling approaches and the derandomisation approach in
~\cite{huang2021efficient}.

\begin{figure}[t]
    \begin{minipage}[t]{0.5\linewidth}
        \centering
        \begin{tikzpicture}[rounded corners] 
            \begin{scope}[local bounding box=graph2, xshift=2.5cm]
            \graph [no placement,
                edge quotes={inner sep=1pt, font=\scriptsize, fill=white},
                nodes={circle, draw, inner sep=3pt,outer sep=0, minimum size=1.5em, text opacity=0, font=\tiny}
            ]
            {
                 0 --[bend right=75, "\(\beta_1(X)\)" left]  1[at=({0em, -3.25em})];
                 0 --["\(\beta_1(Y)\)"]  1;
                 0 --[bend left=75, "\(\beta_1(Z)\)" right]  1;
                 1 --[bend right=75, "\(\beta_2(X)\)" left]  2[at=({0em, -6.8em})];
                 1 --["\(\beta_2(Y)\)"]  2;
                 1 --[bend left=75, "\(\beta_2(Z)\)" right]  2;
                 2 --[bend right=75, "\(\ldots\)" left]  3[at=({0em, -11em})];
                 2 --["\(\ldots\)"]  3;
                 2 --[bend left=75, "\(\ldots\)" right]  3;
                 3 --[bend right=75, "\(\beta_n(X)\)" left]  t[rectangle, sharp corners, at=({0em, -14.25em})];
                 3 --["\(\beta_n(Y)\)"]  t;
                 3 --[bend left=75, "\(\beta_n(Z)\)" right]  t;
            };
            \end{scope}
        \end{tikzpicture}
        \caption{The Decision Diagram of LBCS}
        \label{fig:dd-lbcs}
    \end{minipage}%
    \begin{minipage}[t]{0.5\linewidth}
        \centering
        \begin{tikzpicture}[rounded corners] 
            \begin{scope}[local bounding box=graph2, xshift=2.5cm]
                \graph [no placement,
                    edge quotes={inner sep=2pt, font=\scriptsize, fill=white},
                    nodes={circle, draw, inner sep=3pt,outer sep=0, minimum size=1.5em, text opacity=0, font=\tiny}
                ]
                {
                    0 --["X"] 1[at=({0em, -3.25em})],
                    0 --["Y" ] 5[at=({-6em, -3.25em})],
                    0 --["Z"] 9[at=({6em, -3.25em})],
                    1 --["X"] 12[at=({0em, -6.8em})],
                    12 --["X"] 13[at=({-2em, -11em})],
                    12 -- ["Y"] 14[at=({2em, -11em})],
                    2[at=({-6em, -6.8em})] --["X"] 3[at=({-10em, -11em})],
                    2 --["Y"] 4[at=({-6em, -11em})],
                    3 --["X"] t[rectangle, sharp corners, at=({-2em, -14.25em})],
                    4 --["Y"] t,
                    5 --["Y"] 2,
                    9 --["Z"] 10[at=({6em, -6.8em})],
                    10 --["Z"] 11[at=({6em, -11em})],
                    13 -- ["X"] t,
                    14 -- ["Y"] t,
                    11 --["Z"] t;
                };
            \end{scope}
        \end{tikzpicture}
        \caption{The unoptimised decision diagram of LDF-based Pauli Grouping of H\(_2\) (4 qubits) in Jordan-Wigner encoding}
        \label{fig:dd-ldf}
    \end{minipage}%
\end{figure}

First, it is easy to see that the Decision Diagram (DD) for the LBCS is as in \autoref{fig:dd-lbcs}. Namely, the corresponding DD consists of $n+1$ vertices each of which represents the state after deciding the choice of measuring the previous qubit (excepting the root that represents the start state). For any measurement basis $B = \otimes_{i\in[n]} B_i \in \Pcal^n$ is chosen with probability $\beta(B) = \prod_{i\in[n]}\beta_i(B_i)$ by walking the DD starting from the root and moving along the edges according to the probabilities $\beta_i$ until the terminal. Clearly, we can consider a similar DD for the Pauli grouping in which each path from the root to the leaf in the DD represents the measurement basis of a collection of Pauli operators, as shown in \autoref{fig:dd-ldf} for H$_2$ in Jordan-Wigner encoding. In such case, the Pauli grouping produces YYXX, YYYY, XXXX, XXYY, and ZZZZ each of which is represented as a path of the DD.

\section{Computation and incremental adjustment of zeta-function}
In this part of the appendix we show the computational cost of computing $\zeta(P, \beta)$ for all $P$'s is proportional to the size of the DD thanks to efficient operations of DD. The $\zeta(P, \beta)$ is the probability of choosing a measurement basis $B$ that covers $P$ when random-walking on the DD according to $\beta$. We can compute it by using the \textit{forward weight} and \textit{backward weight} updates of DD in a dynamic-programming manner. Those weight updates were used in~\cite{pmlr-v84-sakaue18a} for efficient sampling from large number of actions. 

For ease of explanation, let us notice that each vertex in the DD can be identified by the sequence of Pauli operators starting from the root to the vertex, say, $P_1 P_2 \ldots P_j$ for $j = 0 \ldots n$ where we define $v()$ as the root. Thus, we write $v(P_1 = X)$ as the vertex reachable from the root by traversing its $X$ edge, and $v(P_1 = X, P_2 = Y)$ as the vertex reachable by another step traversing the $Y$ edge of $v(P_1=X)$, and so on. This allows us to write the probability of choosing $P_{j+1} \in \{X, Y, Z\}$ at $v(P_1\ldots P_j)$ as $\beta(P_{j+1}\mid P_1P_2\ldots P_j)$. Let us define $\zeta(P_{j+1} P_{j+2}\ldots P_n, \beta \mid P_1 P_2 \ldots P_j)$ as the probability of choosing a measurement basis covering $P_{j+1}\ldots P_{n}$ under the condition that the sequence $P_1 \ldots P_j$ has been chosen (i.e., the walk starts from $v(P_1\ldots P_j)$). Clearly, we can see the following relations. At the root, 
\begin{equation*}
    \zeta(P_1P_2\ldots P_n, \beta) =
\begin{cases}
\beta(P_1) \cdot \zeta(P_2\ldots P_n, \beta \mid P_1) &~\text{if}~ P_1 \neq I,\\
\sum_{W \in \{X, Y, Z\}} \beta(W) \cdot \zeta(P_2\ldots P_n, \beta \mid W) &~\text{if}~P_1 = I.  
\end{cases}
\end{equation*}
For $j = 2\ldots n$, the values of $\zeta(P_j\ldots P_n, \beta \mid P_1\ldots P_{j-1})$ can be computed recursively as 
\begin{equation*}
    \zeta(P_j\ldots P_n, \beta \mid P_1 \ldots P_{j-1}) =
\begin{cases}
\beta(P_j \mid P_1 \ldots P_{j-1}) \cdot \zeta(P_{j+1}\ldots P_n, \beta \mid P_1 \ldots P_j) &~\text{if}~ P_j \neq I,\\
\sum_{W \in \{X, Y, Z\}} \beta(W \mid P_1 \ldots P_{j-1}) \cdot \zeta(P_{j+1}\ldots P_n, \beta \mid P_1 \ldots P_{j-1}W) &~\text{if}~P_j = I.  
\end{cases}
\end{equation*}
The above recursive operations allow us to compute $\zeta(P,\beta) = \sum_{B \in \Lift(P)} \beta(B)$ in time linear in the size of DD. 

\section{Choosing the probability distribution}\label{sec:choosing-beta}

How do we choose the probability distribution $\beta$? 
Let us consider the diagonal part of the variance calculation from \autoref{eqn:app-variance-cost} in \hyperref[app:variance]{Appendix~\ref*{app:variance}}. %
\begin{equation}
    \mathrm{cost}_\mathrm{diag}(\beta)
    :=
    \sum_P
    \alpha_P^2~g(P,P,\beta)\Tr(PP\rho).
\end{equation}
This simplifies to
\begin{equation}\label{eq:diagonal-cost}
    \mathrm{cost}_\mathrm{diag}(\beta)
    =
    \sum_P
    \frac{\alpha_P^2}{\zeta(P,\beta)}.
\end{equation}
Can we minimise this cost function in such a way that is compatible with decision diagrams? While for LBCS the cost function can be formulated as a convex optimisation, it is not clear if it remains so with the decision diagram. However, it is quite straightforward to derive a similar iterative procedure as the one used to optimise the probability distribution of the LBCS as described below. 

Notice that at each node of the decision diagram the sum of the probabilities is one, namely, for every $v(P_1\ldots P_{j-1})$ which is a node in the decision diagram it holds that 
\begin{equation}\label{eq:beta-sum-to-one}
    C(v(P_1\ldots P_{j-1})) \equiv \beta\left(X\mid P_1\ldots P_{j-1} \right) + \beta\left(Y\mid P_1\ldots P_{j-1} \right) + \beta\left(Z\mid P_1\ldots P_{j-1} \right) - 1 = 0.
\end{equation}

Combining the two costs of \autoref{eq:diagonal-cost} and \autoref{eq:beta-sum-to-one}, we obtain the following cost function that can be optimised by Lagrange multipliers. 
\begin{equation*}
    L\left(\beta, \mathbf{\lambda} \right) = \sum_P
    \frac{\alpha_P^2}{\zeta(P,\beta)} + \sum_{v~\mbox{is a node of DD}} \lambda_v~C(v). 
\end{equation*}

By taking partial derivatives of the above cost function and by some tedious computation, we derive the following closed-form equation of the (local-optimal) probability distribution. 
\begin{equation}\label{eq:beta-closed-form}
    \beta\left(P_j\mid P_1\ldots P_{j-1} \right) \sim \frac{\sum_Q\left(\alpha_Q\right)^2\sum_{B \in \Lift(Q)~\land~B_j = P_j} \beta(B)}{\left(\zeta(P,\beta)\right)^2}.
\end{equation}

Now, the given the current probability distribution $\beta^{(t)}(B)$ and an update step-size $\Delta \in \left(0, 1\right)$, we can get update the current probability distribution to obtain
\begin{equation*}
    \beta^{(t+1)}(B) = (1-\Delta) \beta^{(t)}(B) + \Delta \beta^{\mbox{closed}}(B),
\end{equation*}
where $\beta^{\mbox{closed}}(\cdot)$ is as in \autoref{eq:beta-closed-form}.

\section{Randomised Pauli Measurements and Their Derandomisation with DDs}\label{app:derandomisation}
We show that DDs may enable us to improve the randomised Pauli measurements and their derandomisation shown in~\cite{huang2021efficient}. First, we restate the problem of Pauli measurements. We are given $L$ Pauli terms $\mathbf{P} = P^{(1)}, P^{(2)}, \ldots, P^{(L)}$. With regards to a fixed (but unknown) quantum state $\rho$, the task is to produce the estimation of $w_l \equiv \Tr(P^{(l)} \rho)$ for $l = 1\ldots L$, denoted as $\hat{w}_l$, by measuring the quantum state $M$ times with $\mathbf{B} = B^{(1)}, B^{(2)}, \ldots, B^{(M)}$ chosen independently at random so that all $w_l$'s are estimated within error $0 < \epsilon \ll 1$ with high confidence (say, with probability $1-\delta$ for $\delta \ll 1$). 

Clearly, we can estimate $w_l$ with the measurement results of all $B^{(m)}$'s that cover $P^{(l)}$ and the number of such $B^{(m)}$'s is $M_l = \sum_{m=1}^M \indicator_{\Lift(P^{(l)})}(B^{(m)})$. The estimation is $\hat{w}_l = \frac{1}{M_l} \sum_{m=1}^M \indicator_{\Lift(P^{(l)})}(B^{(m)}) \cdot \Tr(P^{(l)} \rho)$. Similar to the proof of Lemma~2 in~\cite{huang2021efficient}, we can bound our inconfidence of the estimation as (notice that in~\cite{huang2021efficient} it is called the confidence bound while it is strictly a bound of inconfidence)
\begin{eqnarray*}
\mbox{INCONF}_\epsilon(\mathbf{P}, \mathbf{B}) &\equiv& \Pr\left[\max_{1 \le l \le L} | \hat{w}_l - w_l | 
\ge \epsilon\right] \\
&=& \Pr\left[\cup_{l=1}^L\{ | \hat{w}_l - w_l | \ge \epsilon \} \right] \\
&\le& \sum_{l=1}^L \Pr\left[|\hat{w}_l - w_l | \ge \epsilon  \right] ~~~~~\mbox{(by the union bound)}\\
&\le& 2 \sum_{l=1}^L \exp\left(-\frac{\epsilon^2}{2}\sum_{m=1}^M \indicator_{\Lift(P^{(l)})}(B^{(m)}) \right) ~\mbox{(by the Hoeffding's inequality)}\\
&=& 2 \sum_{l=1}^L \prod_{m=1}^M  \left(1 - \eta\, \indicator_{\Lift(P^{(l)})}(B^{(m)}) \right) ~~~~~\mbox{(by}~\eta \equiv 1 - \exp(-\epsilon^2/2)~\mbox{)}
\end{eqnarray*}

Notice that because
$\EE\left(\indicator_{\Lift(P^{(l)})}(B^{(m)}) \right) = \zeta(P^{(l)}, \beta),$
and due to $B^{(m)}$'s being chosen independently at random, we have
\begin{eqnarray*}
\EE\left( \mbox{INCONF}_\epsilon(\mathbf{P}, \mathbf{B}) \right) &\le& 2 \sum_{l=1}^L 
\left( 
1 - \eta \cdot \zeta(P^{(l)}, \beta)
\right)^M\\
&\le& 2 \sum_{l=1}^L \exp\left(-\eta\cdot \zeta(P^{(l)}, \beta) \cdot M \right)~~\mbox{(due to}~(1-x) \le e^{-x}~\mbox{)}\\
&\le& \delta~~~~~\mbox{(whenever}~M \ge \frac{\ln(2L/\delta)}{\eta}\max_l \frac{1}{\zeta(P^{(l)}, \beta)}~\mbox{)}.
\end{eqnarray*}
Thus, by having $M \propto \frac{\log{(L/\delta)}}{\epsilon^2} \max_{l}\frac{1}{\zeta(P^{(l)}, \beta)}$ we can guarantee with $(1-\delta)$ confidence that all estimations are within $\epsilon$-error. Notice that we have generalised Theorem~1 in~\cite{huang2021efficient} (and Theorem~3 in~\cite{evans2019scalable}) with 
probability distribution $\beta$ obtained from DDs summarized as the following. 
\begin{prop}
Estimators $\hat{w}_l$ for $\mathbf{P}$ obtained with measurements $\mathbf{B}$ sampled from a compatible probability distribution $\beta$ of a decision diagram result in $\epsilon$-error predictions of $w_l = \Tr(P^{(l)}\rho)$ for $l = 1\ldots L$ up to an additive error whenever $M \propto \frac{\log{L}}{\epsilon^2} \max_{l}\frac{1}{\zeta(P^{(l)}, \beta)}$. 
\end{prop}

Going further, we should be able to derandomise the measurement bases by similar procedures. For example, by sampling every path (or, measurement basis) in the DD uniformly at random, $\zeta(P, \beta)$ is the ratio of the number of paths covering $P$ against the total number of paths, which can be larger than $1/3^{|\supp(P)|}$. A similar argument can be used to derandomise the LBCS to obtain better estimators. We leave the details as future work.

\end{document}